# A negative index metamaterial driven by phonons on a ZnO platform

**Julia Inglés-Cerrillo[1], Pablo Ibañez-Romero[1], Rajveer Fandan[1], Jorge Pedrós[1], Nolwenn Le Biavan[2], Denis Lefebvre[2], Maxime Hugues[2], Jean-Michel Chauveau[2,3], Miguel Montes Bajo[1], Adrian Hierro[1]**

[1]ISOM, Universidad Politécnica de Madrid, Spain
[2]Université Côte d'Azur, CNRS, CRHEA, Valbonne, France
[3]Université Paris Saclay, Université Versailles Saint Quentin, CNRS, GEMaC, Versailles France

**Contact authors:** miguel.montes@upm.es, adrian.hierro@upm.es

**ABSTRACT**

Negative index metamaterials (NIMs) can be achieved with uniaxial hyperbolic metamaterials (HMMs) featuring $\epsilon_\parallel > 0$ and $\epsilon_\perp < 0$. This type of approach has been traditionally realized using stacked doped/undoped semiconductor layers. Only recently surface phonon polaritons (SPhPs) have emerged as a promising low-loss alternative to surface plasmon polaritons (SPPs). Despite this advantage, the SPhP-based approach has been underexplored due to the challenges associated with ensuring high crystal quality in the heterostructure when using alloys with different phonon frequencies. In this work, we design a phononic-driven NIM using a ZnO/(Zn,Mg)O heterostructure, demonstrating control over its hyperbolic behavior through the precise selection of the Mg content and the relative layer thicknesses. Our study shows that increasing the Mg content in the ternary layers enhances the type I behavior, and that the optimal layer thickness varies depending on the Mg content. After analyzing the conditions for achieving type I hyperbolic dispersion, we experimentally demonstrate this concept with a sample featuring equal layer thicknesses and a 32% Mg concentration. We characterize the structure by means of polarized reflectance spectroscopy and use attenuated total reflectance spectroscopy to report the presence of a SPhP mode located within the type I hyperbolic region. By employing the transfer matrix method, we demonstrate that this mode exhibits negative frequency dispersion, a hallmark of type I hyperbolic modes, and isofrequency curve calculations further confirm this behavior. Controlling the design of a phononic hyperbolic type I metamaterial lays the groundwork for exploring its potential applications in attaining low-loss, sub-diffraction-limited optical modes using SPhP excitations.

**INTRODUCTION**

An optical metamaterial is an artificial material with rationally designed properties in which the macroscopic electromagnetic properties are not dictated by the individual materials that compose it, but by the sub-wavelength features of the entire structure.[1] Possibly, the most important group of metamaterials are hyperbolic metamaterials (HMMs)[2], defined as uniaxial materials presenting an anisotropic dielectric function in which the real part of the permittivity has opposite signs in perpendicular directions. This anisotropy enables HMMs to exhibit unique properties, such as strong enhancement of spontaneous emission, diverging density of states, enhanced superlensing effects, or a negative refractive index.[3,4,5,6] Because of these distinctive properties, metamaterials are expected to open a new realm of functionalities unattainable from naturally occurring materials. One of the most exciting opportunities that HMMs offer is the development of negative-index materials (NIMs).[4,7]

Negative phase velocity and its implications were first explored by Sir Arthur Schuster[8] and H.Lamb[9] as early as 1904. After this, it was Veselago[10] who provided the modern description of "negative permittivity/negative permeability" for negative refraction. The surge of interest in NIMs over the past decades was driven by Sir John Pendry, whose contributions to the field include a prediction of the NIM-based superlens capable of achieving resolution beyond the diffraction limit.[11] Since then, various approaches have been explored to achieve NIMs. The first approach consisted on fabricating a metamaterial which exhibited both negative electric permittivity ($\epsilon < 0$) and negative magnetic permeability (μ < 0), achieved by the arrangement of materials in complex periodic nanostructures.[12,13] However, the complexity of the fabrication process led to the development of an alternative approach, which consisted on stacking layers of different materials to achieve positive/negative components of the dielectric tensor along and perpendicular to the growth plane, respectively. The resulting metamaterial behaves as a metal in the out-of-plane direction and as a dielectric in the in-plane one. In this case, one needs to consider the dispersion relation for uniaxial anisotropic materials, given by:

$$\frac{k_\parallel^2}{\epsilon_\perp} + \frac{k_\perp^2}{\epsilon_\parallel} = \frac{\omega^2}{c^2} \qquad (1)$$

where the subscripts ∥ and ⊥ indicate the directions parallel and perpendicular to the surface, respectively. The solutions to **Equation (1)** are hyperboloids of either type I ($\epsilon_\parallel > 0$ and $\epsilon_\perp < 0$) or type II ($\epsilon_\parallel < 0$ and $\epsilon_\perp > 0$).[5,6,2] Analyzing the corresponding isofrequency curves, it can be seen that type I hyperbolic modes exhibit negative dispersion ($\frac{d\omega}{dk_\parallel} < 0$), while type II modes present positive dispersion ($\frac{d\omega}{dk_\parallel} > 0$).[5,14,15] Hence, the sign of the mode dispersion can be used to confirm the hyperbolic character of the metamaterial in a specific spectral region.

Hyperbolic metamaterials can be achieved by stacking materials with opposite sign of the dielectric function, which can be controlled via a resonance (i.e. a zero of the dielectric function) in the energy region of interest. This resonance can be easily obtained with a plasmon, and hence by growing doped/undoped semiconductor layers forming a heterostructure the metamaterial can become hyperbolic type I. The first demonstration of a NIM using this approach was published by Hoffman et al.[4] where they achieved a comparatively low-loss, all-semiconductor metamaterial exhibiting negative refraction in the infrared region, by using alternating layers of highly doped InGaAs and intrinsic AlInAs. Since then, this plasmonic approach has been widely used as the preferred method for obtaining hyperbolic metamaterials.[16,2,17] Good examples of type I hyperbolic metamaterials have been demonstrated in doped/undoped ZnO stacked layers,[5] or in ZnO/(Zn,Mg)O multiple quantum wells with the added functionality of quantum control of the optical behavior.[18]

However, the metallic out-of-plane behavior of the anisotropic metamaterial can also be obtained by substituting the plasma resonance with other resonances also present in crystals, like phonon resonances. Within this group, surface phonon polariton (SPhP) modes have emerged as a compelling area of research on their own right, owing to their significantly higher quality factor compared to surface plasmon polaritons (SPPs),[19] along with the potential for tunability achieved through precise control and design of nanostructure geometry, size and periodicity.[19] For example, Caldwell et al., demonstrated that lithographic fabrication could be used to design SPhP resonators with *a priori* designed resonant frequencies, using 6H-silicon carbide nanopillar antenna arrays.[20] In addition, efforts have been made to manipulate SPhP modes at the nanoscale. Huber et al. accomplished the focusing of propagating SPhP modes on a SiC surface using a straightforward concave gold focusing element,[21] demonstrating sub-diffraction focusing, which is highly relevant for nanophotonic applications.

Due to their vast potential and wide range of applications, phonon resonances have reached the field of metamaterials, but only very recently. It was not until five years ago, in 2019, that Daniel C. Rachford et al. demonstrated the first phononic hyperbolic metamaterial, using AlN/GaN superlattices supporting SPhPs.[14] With the phononic approach the difficulty relies on the fact that to obtain the desired hyperbolic behavior one needs to use alloys with different phonon frequencies, but with the capability to be stacked together without degrading the overall quality of the heterostructure. Even though achieving the desired metallic/dielectric behavior can be more challenging with phonons than plasmons, SPhPs present substantially reduced losses in comparison to SPPs, related to the longer scattering times of optical phonons compared to the fast scattering times of electrons in metals and semiconductors. This yields a quality factor for localized modes significantly higher in SPhP materials than in plasmonic materials.[3,19] For this reason, phonon modes appear promising with respect to plasmonic modes from the viewpoint of realizing low-loss, sub-diffraction-limited optical modes, hence our focus on this last approach.

In this work, we set the design rules of a NIM driven by phonon resonances to achieve the maximum figure of merit, and we provide an experimental demonstration of the concept. To do so, we build and analyze a uniaxial type I hyperbolic heterostructure composed of layers of highly polar ZnO/(Zn,Mg)O alloys whose LO phonon frequencies can be controlled with the alloy composition. Secondly, we select one case that meets the requirements to exhibit negative refraction, and experimentally demonstrate the existence of type I hyperbolic phonon polariton modes.

**RESULTS AND DISCUSSION**

**TUNING OF HYPERBOLIC DISPERSION WITH PHONONS**

In order to generate the desired stack of layers with alternating signs of the dielectric function, controlled by phonon resonances, we take advantage of the shift of the LO phonon frequency (i.e. the zero-crossing point of the real part of the dielectric function) of wurtzite (Zn,Mg)O from 590 to 690 cm$^{-1}$ with increasing Mg content.[22,23] **Figure 1(a)** shows the real part of the dielectric function of alloys with 0, 10 and 40% of Mg. It can be observed that as we increase the Mg concentration, the $\omega_{LO2}$ phonon frequency of the ternary material also increases, while all the other phonon frequencies do not shift significantly with [Mg]. As can be seen in the inset of **Figure 1(a)**, there are frequency regions over which ZnO behaves as a dielectric whereas (Zn,Mg)O shows a metallic behavior.

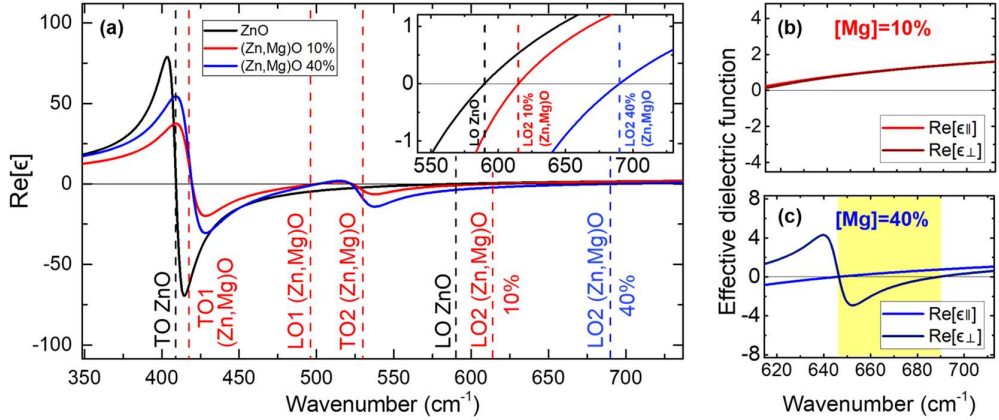

**Figure 1:** (a) Dielectric functions of the constituent materials of the heterostructure ($\epsilon_{ZnO}$ and $\epsilon_{(Zn,Mg)O}$). Reference lines are included as a guide to locate each phonon frequency. The inset focuses on the region where the zero-crossing point of the real part of the dielectric function is found, showing that this point can be shifted with the Mg content. (b), (c) In-plane ($\epsilon_\parallel$) and out-of-plane ($\epsilon_\perp$) components of the effective dielectric function of the metamaterial for $d_{ZnO}/d_{(Zn,Mg)O} = 1$, across the spectral region where the hyperbolic behavior can be achieved. The type I HMM behavior is highlighted in yellow, and can be achieved for 40% of [Mg] but not for 10%.

As a first step, we need to define and analyze the permittivity functions of the two constituent materials separately. In the case of the (Zn,Mg)O alloy, the response to light can be described with a Gervais oscillator modelling the interaction of the light with two pairs of phonon frequencies denoted as TO$_1$-LO$_1$ and TO$_2$-LO$_2$, while in the case of ZnO there is only one pair of TO-LO phonons. The expressions defining the two permittivity functions can therefore be written as follows[24]:

$$\epsilon_{(Zn,Mg)O}(\omega) = \epsilon_\infty^{(Zn,Mg)O} \prod_{k=1}^{2} \frac{\omega_{LO,k}^2 - \omega^2 - i\omega\gamma_{LO,k}\omega}{\omega_{TO,k}^2 - \omega^2 - i\omega\gamma_{TO,k}\omega} \qquad (2)$$

$$\epsilon_{ZnO}(\omega) = \epsilon_\infty^{ZnO} \frac{\omega_{LO}^2 - \omega^2 - i\omega\gamma_{LO}}{\omega_{TO}^2 - \omega^2 - i\omega\gamma_{TO}} \tag{3}$$

where the parameter $\epsilon_\infty$ is the high frequency dielectric constant, $\omega_{LO}$ and $\omega_{TO}$ are the longitudinal and transverse optic phonon frequencies, and $\gamma_{LO}$ and $\gamma_{TO}$ their respective damping parameters. The values for the phonon frequencies and damping parameters are taken from our previous work on similar heterostructures.[23,24]

Once the permittivity functions of the individual materials are defined, the effective medium theory approximation is employed to describe the effective anisotropic dielectric function of the ZnO/(Zn,Mg)O metamaterial. The two components of the effective dielectric function, which define the optical properties of the metamaterial in the in-plane and out-of-plane directions, are defined as follows:

$$\epsilon_\parallel = \frac{d_{ZnO}\epsilon_{ZnO} + d_{(Zn,Mg)O}\epsilon_{(Zn,Mg)O}}{d_{ZnO} + d_{(Zn,Mg)O}} \tag{4}$$

$$\epsilon_\perp = \left(\frac{\frac{d_{ZnO}}{\epsilon_{ZnO}} + \frac{d_{(Zn,Mg)O}}{\epsilon_{(Zn,Mg)O}}}{d_{ZnO} + d_{(Zn,Mg)O}}\right)^{-1} \tag{5}$$

where $\epsilon_{ZnO}$ and $\epsilon_{(Zn,Mg)O}$ are the dielectric functions of the two constituent materials, and $d_{ZnO}, d_{(Zn,Mg)O}$, their respective layer thicknesses. The real part of the two components of the effective dielectric function of the metamaterial are shown in **Figure 1(b),(c)** for two representative cases, 10% and 40% of [Mg], and a thickness ratio of 1. It can be seen that, focusing on the phononic region of interest, a type I hyperbolic region is achieved for 40% of [Mg] but not for 10%.

To study the type I HMM behavior in ZnO/(Zn,Mg)O heterostructures, the in-plane and out-of-plane components of the dielectric function are analyzed to determine if there exists a spectral range where $\epsilon_\parallel > 0$ and $\epsilon_\perp < 0$. Two parameters are varied for this analysis: the Mg concentration in the (Zn,Mg)O layers, and the thickness ratio between layers ($d_{ZnO}/d_{(Zn,Mg)O}$). In order to support that the hyperbolic mode is indeed type I, the negative slope sign of the moe dispersion is confirmed by employing the transfer matrix method to evaluate the imaginary part of the reflection coefficient ($r_p$) for high values of $k_\parallel$, as well as by performing isofrequency curves calculations. Finally, the quality of the hyperbolic mode is quantified through a figure of merit given by $FOM = Re(k_\perp)/Im(k_\perp)$, typically accepted for characterizing anisotropic negative index materials.[4] Details on the calculations for the mode dispersion and FOM are given in the Methods Section.

**Figure 2** illustrates the role played by the Mg concentration in the hyperbolic behavior in a ZnO/(Zn,Mg)O metamaterial with a thickness ratio of $d_{ZnO}/d_{(Zn,Mg)O}$ = 1. For the calculation, we account for the highest Mg concentration that can be achieved in wurtzite (Zn,Mg)O grown on a ZnO substrate before relaxation, which depends on the (Zn,Mg)O total thickness, and can be as high as 40%.[25,26] The panels in the left column of **Figure 2** show the calculated effective dielectric function in the phononic spectral region for four Mg concentrations (10%, 20%, 30% and 40%) in the ternary layers. The type I HMM region is highlighted in yellow. It can be observed that a minimum Mg concentration is needed for the type I hyperbolic region to exist, and once the type I region appears, it gets broader with increasing [Mg]. This is due to the fact that increasing the Mg concentration increases the $\omega_{LO2}$ phonon frequency of the (Zn,Mg)O layer[22], which leads to a larger difference between the $\omega_{LO}$ phonon frequencies of the ZnO and (Zn,Mg)O layers, thereby broadening the type I hyperbolic region.

For each of these values of [Mg], we calculated the corresponding absorption coefficient and figure of merit (FOM) of the metamaterial, shown in the middle column of **Figure 2**. In the case of [Mg]=10%, the FOM has no physical meaning, as a type I hyperbolic region does not exist, whereas in the cases of 20%, 30% and 40%, increasing the Mg concentration yields an increasing value of the FOM. If we now look at the wavenumbers where the FOM reaches a maximum value, it can be observed that the absorption coefficient at these wavenumbers decreases as the Mg concentration increases, due to the shift of the $\omega_{LO2}$ phonon frequency with [Mg]. Furthermore, with increasing [Mg], the peak value of the FOM shifts away from the maximum value of the absorption coefficient. All in all, by introducing higher Mg we achieve a less absorbent, better-quality metamaterial.

Finally, from the dispersion relation for the four different values of [Mg] shown in the panels in the right column of **Figure 2,** it can be seen that the type I hyperbolic mode presents a negative slope. This, as discussed in the introduction, is a characteristic feature of type I hyperbolic modes. Furthermore, with increasing [Mg] this mode becomes more dispersive, and the maximum value reached by the imaginary part of the Fresnel reflection coefficient for p-polarized light ($\text{Im}[r_p]$) increases.

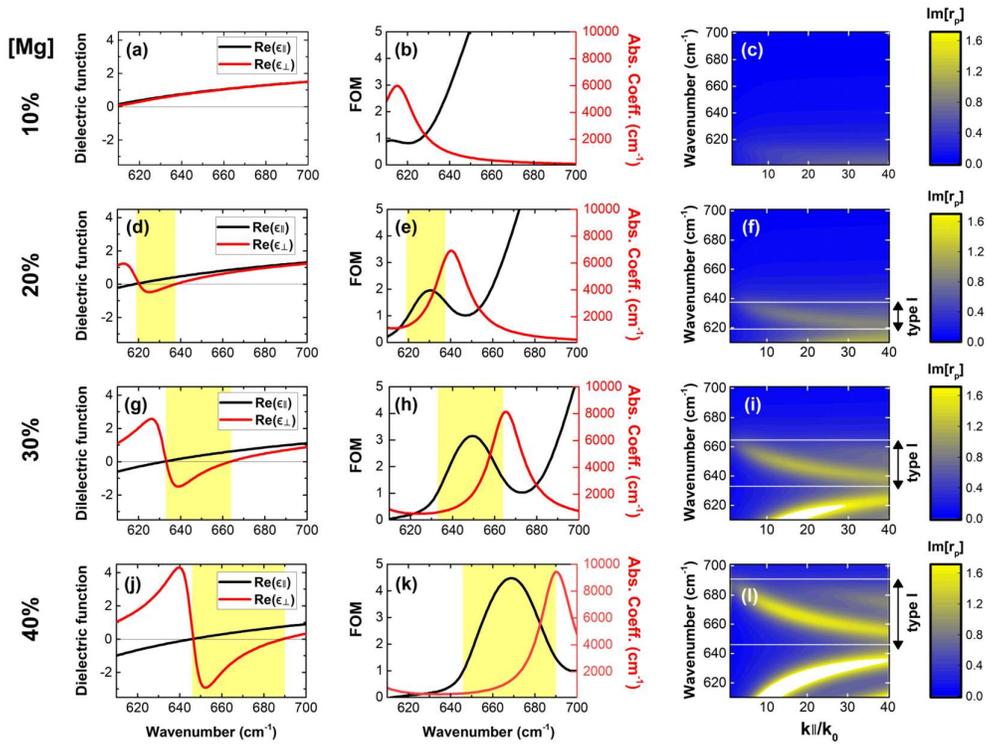

**Figure 2: Type I hyperbolic region dependence on [Mg] for a thickness ratio $d_{ZnO}/d_{(Zn,Mg)O}$ = 1. For Mg concentrations of 10%, 20%, 30% and 40%, the panels in the left column show the real part of the calculated effective dielectric function of the metamaterial, and the panels in the middle column show the figure of merit (FOM) and absorption coefficient of the metamaterial. In these first two columns, the type I HMM behavior is highlighted in yellow. Panels in the right column show the calculated dispersion curves of the metamaterial, where the color legend corresponds to the imaginary part of the Fresnel reflection coefficient for p-polarized light, $\text{Im}[r_p]$, as a function of the relative in-plane momentum $k_\parallel/k_0$, where $k_0$ is the wavevector in vacuum.**

To further confirm the nature of each mode observed in the dispersion relations, we select the case with [Mg]=30% and calculate the isofrequency curves for three different frequencies: one below (620 cm$^{-1}$), one within (650 cm$^{-1}$), and one above (680 cm$^{-1}$) the type I hyperbolic region. As anticipated, the obtained isofrequency curves (IFCs) exhibit a hyperbolic shape characteristic of the type II hyperbolic region, a different hyperbola characteristic of the type I hyperbolic region, and an ellipse corresponding to the dielectric region, for the three frequencies chosen, respectively. The three IFCs are shown in **Figure S1** of the Supporting Information.

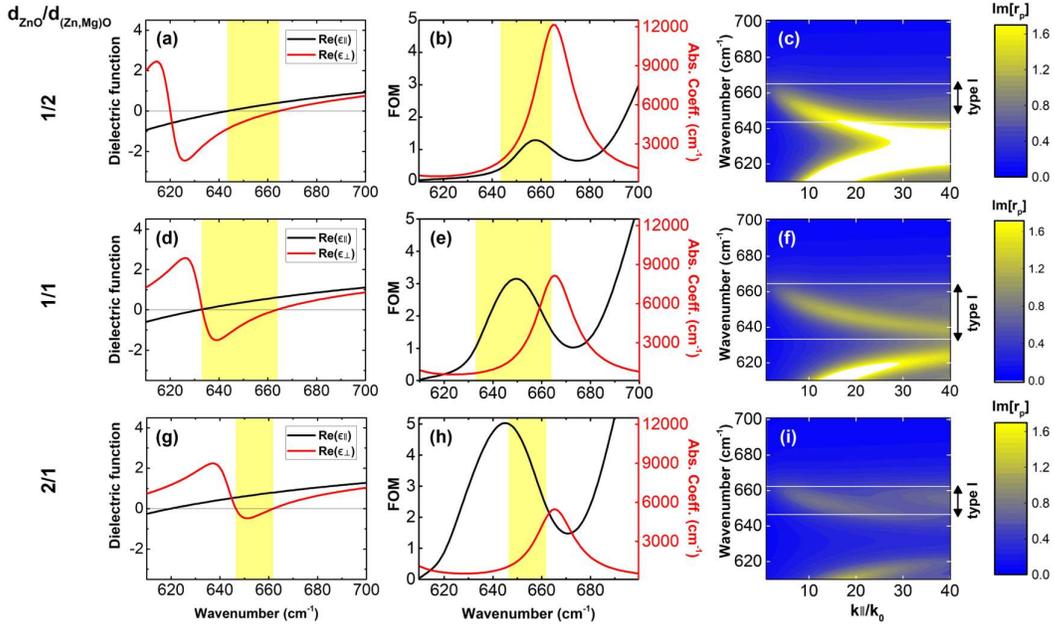

**Figure 3:** Type I hyperbolic region dependence on thickness ratio for [Mg]=30%. For $d_{ZnO}/d_{(Zn,Mg)O}$ of 1/2, 1/1 and 2/1, the panels in the left column show the real part of the calculated effective dielectric function of the metamaterial, and the panels in the middle column show the figure of merit (FOM) and absorption coefficient of the metamaterial. In these two columns, the type I HMM behavior is highlighted in yellow. Panels in the right column show the calculated dispersion curves of the metamaterial, where the color legend corresponds to the Imaginary part of the Fresnel reflection coefficient for p-polarized light $\mathrm{Im}[r_p]$ as a function of the relative in-plane momentum $k_\parallel/k_0$, where $k_0$ is the wavevector in vacuum.

Analogously to the analysis of the Mg content, the impact of the layer thickness ratio on the hyperbolic behavior is shown in **Figure 3**, for a Mg content of 30% and three representative thickness ratios: $d_{ZnO}/d_{(Zn,Mg)O}$ of 1/1 (when both layers have the same thickness), 1/2 and 2/1 (when the (Zn,Mg)O layer is twice the thickness of the ZnO layer, and vice versa).

To discuss the different cases, we will first analyze the behavior of the effective dielectric function in this region, shown in the left column of **Figure 3**. When both layers have the same thickness, **Figure 3(d)** shows that the low frequency limit occurs at the frequency where the in- and out-of-plane components of the effective dielectric function simultaneously change sign, and the upper frequency limit when the out-of-plane component does. Below the low frequency limit, around 633 cm$^{-1}$, we find a direct transition from a type I to a type II hyperbolic region ($\epsilon_\parallel < 0$ and $\epsilon_\perp > 0$). This direct transition between the two different types of hyperbolic regions at equal layer thicknesses is in agreement with Ref.[17]. However, when the relative layer thicknesses change, this direct transition disappears, and we find the emergence of either an effective dielectric or an effective metallic region between the type I and the type II hyperbolic regions.

If $d_{ZnO} < d_{(Zn,Mg)O}$, as is the case of **Figure 3(a)**, the Lorentzian line shape of $\epsilon_\perp$ redshifts while the zero crossing of $\epsilon_\parallel$ blueshifts, resulting in the emergence of a region below the low frequency limit in which both the in- and out-of-plane components of the dielectric function are negative, ie. an effective metallic region. **Figure S2** in the Supporting Information shows that the type I hyperbolic region gets narrower if the thickness ratio is further decreased. In the opposite case, if $d_{ZnO} > d_{(Zn,Mg)O}$, it can be observed in **Figure 3(g)** that the minima of $\epsilon_\perp$ moves upwards, which results in the emergence of an effective dielectric region below the type I hyperbolic region. In this case, **Figure S2** in the Supporting Information shows that when the thickness ratio is further increased, $\epsilon_\perp$ is no longer negative in the phononic range of interest and the material becomes dielectric in this region.

These three different cases influence the behavior of the dispersion of the type I hyperbolic mode. When both layers have the same thickness, **Figure 3(f)** shows that the type I mode presents a negative slope, as expected, and that the dispersion of this mode is delimited by the boundaries of the type I hyperbolic region (marked by the condition $\epsilon_\parallel > 0$ and $\epsilon_\perp < 0$). However, when $d_{ZnO} < d_{(Zn,Mg)O}$ an effective metallic region emerges adjacent to the hyperbolic type I one, as is the case in **Figure 3(c)** in which the metallic region is found from 613 to 649 $cm^{-1}$, and the dispersion of the mode spans the entire type I region and enters the metallic region. Finally, when $d_{ZnO} > d_{(Zn,Mg)O}$, **Figure 3(i)** shows that the hyperbolic mode loses intensity.

In the panels in the middle column, it can be seen that as the (Zn,Mg)O layer becomes thicker than the ZnO layer (**Figure 3(b)**), the FOM decreases and its peak value shifts closer to the absorption maximum value. This implies that the type I hyperbolic region is more absorbent, which is undesirable in terms of light manipulation purposes. On the other hand, if the ZnO layer becomes thicker than the (Zn,Mg)O layer, the FOM peak value shifts away from the absorption coefficient maximum value, as can be seen in **Figure 3(h)**. This behavior is analogous to the one found when the Mg concentration is increased, which is beneficial from an applicability point of view.

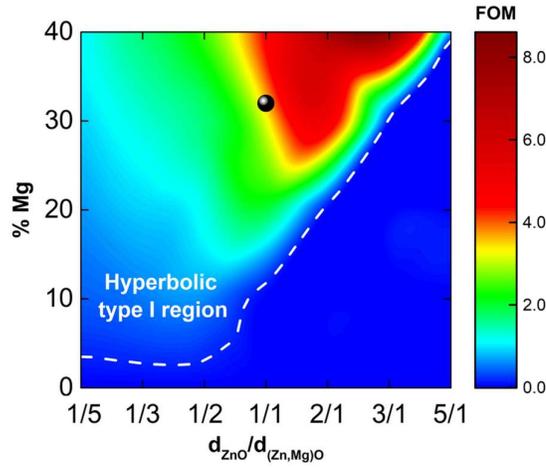

Figure 4: Color map showing the peak value of the FOM ($Re(k_\perp)/Im(k_\perp)$) as a function of [Mg] and the thickness ratio. In the areas of the map where there is no type I hyperbolic region, the FOM is set to zero. The white dashed line defines the boundary between the region of parameter combinations for which a type I region is found and those for which it is not. The black dot indicates the experimental sample used in this work.

To further extend the analysis of **Figures 2 and 3** to a larger number of design conditions, **Figure 4** shows the peak value of the FOM in the spectral region where the condition for HMM type I is met, for different combinations of the two parameters, [Mg] and $d_{ZnO}/d_{(Zn,Mg)O}$. The [Mg] values are limited within 0 and 40% to narrow the study as to explore only experimentally feasible ZnO/Zn(Mg)O heterostructures. The thickness ratio is confined to the range 1/5 to 5/1 because beyond these limits, as the thicknesses of the two layers differ more from each other, the hyperbolic behavior is lost. In the combination of [Mg] and thickness ratio where there is no HMM type I, the FOM is set to zero. This more detailed study shows that the type I HMM behavior is enhanced when a high [Mg] is incorporated into the ternary layers, and that the optimal relative layer thickness to achieve the highest value of the FOM is not always the same but varies depending on the Mg content. Generally, as the Mg content increases, the maximum FOM is achieved with larger $d_{ZnO}/d_{(Zn,Mg)O}$ thickness ratios. For example, for a [Mg] value of 40%, the highest FOM would be achieved for a value of $d_{ZnO}/d_{(Zn,Mg)O}$ between 2/1 and 3/1, while for a lower [Mg] value, such as 10%, the highest FOM would be achieved with $d_{ZnO}/d_{(Zn,Mg)O}$ between 1/3 and 1/2.

It is worth mentioning that heterostructures similar to those presented in this work are usually found in the form of doped-multiple quantum wells within the active region of various devices (in the case of the (Zn,Mg)O/ZnO platform, for example Ref.[27]). In this case, one needs to account for the interaction of light with free electrons oscillating at the plasma frequency, effect that needs to be added to **Equation (3)**. Then the question is whether these doped structures can be tuned to also become HMM type I in the phononic region. However, as shown in **Figure S3** of the Supporting Information, the plasma resonance causes the entire ZnO dielectric function to become negative within the range of interest, making the multilayer structure behave as a metallic material.

**DEMONSTRATION OF A TYPE I HYPERBOLIC PHONONIC METAMATERIAL**

Having analyzed the necessary conditions to obtain a material exhibiting type I hyperbolic dispersion with ZnO/(Zn,Mg)O heterostructures, we now experimentally demonstrate this concept on a sample with a Mg concentration of 32% and a thickness ratio of $d_{ZnO}/d_{(Zn,Mg)O}$ =1/1, with thicknesses of 5 nm each, and grown on a ZnO substrate. This sample is indicated in **Figure 4** with a black dot symbol and, considering its structure (see **Figure 5(a)**), it should display HMM type I character. Details on the growth conditions are given in the Methods Section.

We use polarized reflectance spectroscopy to characterize the structure. Fitting the reflectance spectra with a transfer matrix method (TMM) that accounts for the individual layers described by the dielectric function expressions from **Equation (2) and (3)** allows us to obtain the TO-LO frequencies and losses of the ZnO and (Zn,Mg)O layers, their thicknesses, and the Mg content in the sample (see Ref.[23] for the determination of Mg). Details on these values are given in **Table S1** of the Supporting Information. As shown in **Figure 5(b)**, the reflectance calculated with the TMM model (black dotted lines) resembles very well the measured spectra (solid blue lines), especially in the region where the phonon related features are found (**Figure 5(c)**). With the parameters obtained from the fit, the effective dielectric function is calculated using **Equation (4) and (5)**. The validity of the calculated effective dielectric function of the metamaterial is confirmed by calculating the reflectance spectrum of the structure, substituting the 40 alternating ZnO/(Zn,Mg)O individual layers by a unique layer of the same total thickness defined by an effective medium. Hence, the red dotted lines in **Figure 5(b),(c)** show that this reflectance agrees quite well with the measured spectrum and with that calculated accounting for the individual layers. The values obtained for $\epsilon_\parallel$ and $\epsilon_\perp$ are plotted in **Figure 5(d)**, indicating the presence of a type I hyperbolic behavior of the sample in the spectral region between 635 and 668 cm$^{-1}$. Finally, **Figure 5(e)** shows the calculated FOM and absorption coefficient for the structure, where a value of around 3 is obtained for the FOM, comparing well with the state of the art.[4]

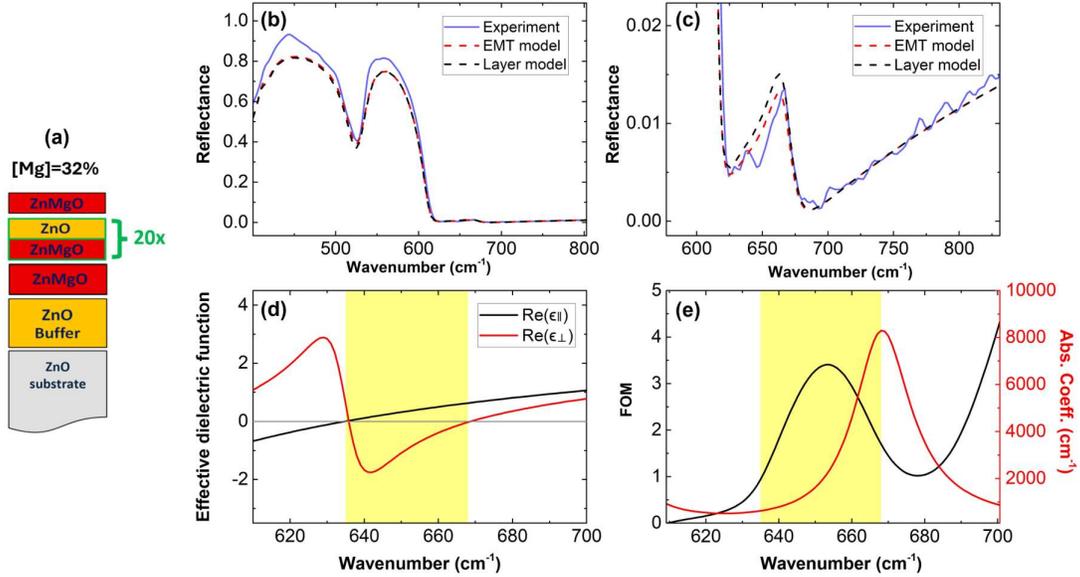

**Figure 5:** (a) Illustration depicting the measured sample. The layers marked in green correspond to the ZnO/(Zn,Mg)O metamaterial, consisting of 40 alternating layers (i.e., 20 repetitions), with each layer having an equal thickness of 5 nm. (b) Reflectance spectra under p-polarization at 45° angle of incidence. The solid blue line is the experimental result, the dotted black line is the fitted reflectance spectra using the TMM model accounting for all individual layers, and the dotted red line is the calculated reflectance spectra using the effective medium theory (EMT) approximation. (c) Focus on the phononic region of interest. (d) Effective dielectric function calculated using the values of phonon frequencies and dampings extracted from fitting the layer model to the experimental results. The type I HMM behavior is highlighted in yellow. (e) Figure of merit (FOM) and absorption coefficient of the measured sample.

One key test for validating the experimentally extracted effective dielectric function of **Figure 5(d)** is to make accurate predictions of the polariton modes present in the structure. To be able to experimentally observe SPhPs, reflectance ATR measurements were carried out using a ZnSe prism in the Otto configuration (see Methods Section for details). **Figure 6(a)** shows the measured ATR curves for four different angles of incidence. For all angles of incidence, there is a SPhP which is clearly visible as a minimum in the reflectance curves, at a frequency that lies within the type I hyperbolic spectral range (635 – 668 cm$^{-1}$). The ATR reflectance curves are also simulated with a TMM that accounts for the effective medium, as a function of the angle of incidence, resulting in the contour plot in **Figure 6(b)**, where the experimental SPhP values have also been plotted with black dots. It can be observed that the modelled branch corresponding to the SPhP, which lies within the type I hyperbolic region, is in excellent agreement with the positions of the reflectance minima extracted from the ATR measurements in **Figure 6(a)**. Having observed experimentally the presence of the SPhP within the type I hyperbolic region, the type I character of this mode can be corroborated by analyzing its frequency dispersion. For this purpose, dispersion calculations were performed for high values of $k_\parallel$ using the TMM for an anisotropic material and evaluating the imaginary part of the reflection coefficient $r_p$, which peaks at polariton resonances[28] (see Methods Section). **Figure 6(c)** displays the resulting dispersion curves of our sample, revealing various modes. As expected, negative dispersion indicative of type I hyperbolic mode progression is observed on the spectral region where we anticipated observing type I HMM behavior (635 – 668 cm$^{-1}$). Finally, we calculate the isofrequency curve for the measured structure at the frequency where the position of the reflectance minima corresponding to the SPhP mode is experimentally observed (663 cm$^{-1}$). As illustrated in **Figure 6(d)**, the resulting curve is a hyperbola characteristic of type I hyperbolic region, providing definitive evidence of the HMM type I behavior. Alternatively, the isofrequency curve can also be calculated at the frequency where the peak value of the FOM is found for this structure (653 cm$^{-1}$). This is shown in **Figure S4** of the Supporting Information.

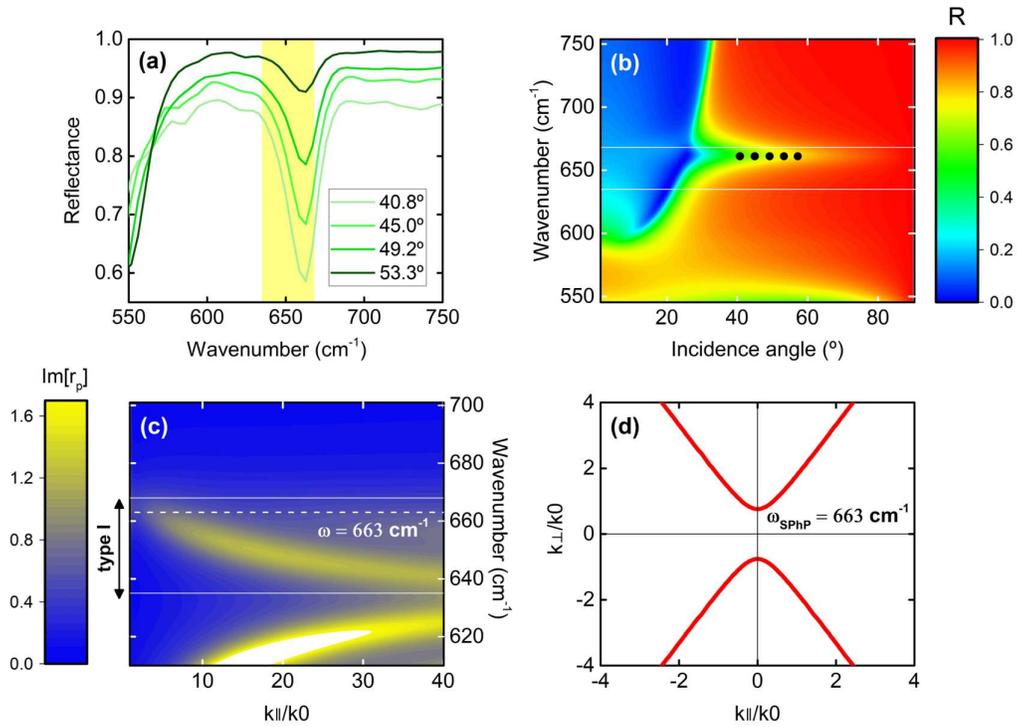

**Figure 6: (a)** Measured attenuated total reflectance curves in p-polarization at different angles of incidence. The type I HMM behavior is highlighted in yellow. **(b)** Simulated ATR contour plot and experimentally determined resonance frequencies (circular black dots) of the SPhP mode for the four different angles of incidence shown in (a). The white horizontal lines indicate the hyperbolic region type I corresponding to the measured sample. **(c)** Calculated polariton dispersion curves of the measured sample, computed with the TMM. Solid white reference lines show the spectral region where the type I behavior has been found (635 – 668 cm$^{-1}$). The color legend corresponds to the Imaginary part of the Fresnel reflection coefficient for p-polarized light $\mathrm{Im}[r_p]$ as a function of the relative in-plane momentum $k_\parallel/k_0$, where $k_0$ is the wavevector in vacuum. **(d)** Calculated isofrequency curve of the measured sample for a frequency of 663 cm$^{-1}$, where the SPhP excitation is measured, and which lies within the type I hyperbolic region. This frequency is marked in (c) with a white dashed line. As expected, the IFC shows a hyperbolic shape characteristic of type I hyperbolic region. $k_\parallel$ and $k_\perp$ are the parallel and perpendicular components of the wavevector, respectively. Both variables are normalized to $k_0$, which is the wavevector in vacuum.

## CONCLUSIONS

In summary, we have performed a detailed analysis of the design rules to obtain a phononic type I hyperbolic metamaterial using a realistic ZnO/(Zn,Mg)O heterostructure. The concept here is to control the epsilon-near-zero character of the alternating layers by tuning their phononic resonance. This is achieved by shifting the LO phonon frequency via the Mg content in the ternary alloy, but accounting for the realistic values of the phonon dampings and the maximum achievable Mg content in a heterostructure. By using a high Mg content, and selecting the appropriate relative thickness value of the ZnO and (Zn,Mg)O layers, a phononic-driven type I hyperbolic region can be obtained with a FOM comparing quite well with the state of the art. This concept can be generalized to other polar semiconductor heterostructures where the phonon frequency is a function of the alloy composition.

Experimentally, we have demonstrated this concept in a ZnO/(Zn,Mg)O heterostructure with a thickness ratio of 1/1, and a Mg concentration of 32%. In this sample, we observe a hyperbolic type I behavior between 635 and 668 cm$^{-1}$, region defined by the ZnO and (Zn,Mg)O LO phonon frequencies. By using ATR spectroscopy and a TMM model, we have demonstrated the presence of a surface phonon polariton within this spectral region, which, as expected, shows a negative frequency dispersion, characteristic of hyperbolic type I modes. The design and demonstration of a phononic hyperbolic type I metamaterial through the precise selection of parameters in polar heterostructures paves the way for exploring its potential applications in attaining low-loss, sub-diffraction-limited optical modes using SPhP excitations.

**METHODS**

**Sample growth.**

The ZnO/(Zn,Mg)O heterostructure employed in this work was grown by molecular beam epitaxy (MBE) on a m-plane ZnO substrate as shown in **Figure 5(a)**. The structure consists of 40 alternating layers of ZnO and (Zn,Mg)O, with a thickness of 5 nm each, atop a 100-nm thick ZnO buffer layer grown on the ZnO substrate. The (Zn,Mg)O layers have a nominal Mg concentration of 32%, thus presenting a wurtzite structure.[25,26]

**Polarized reflectance spectroscopy measurements and ATR measurements.**

Reflectance measurements were carried out in a Fourier Transform Infrared (FTIR) Spectrometer, using s- and p-polarized incident light. Given its wurtzite structure, in the ZnO/(Zn,Mg)O heterostructures twelve phonon branches exist.[22] For the lattice vibrations with $E_1$ symmetry, the atoms move perpendicular to the c-axis.[22] In order to simplify our analysis of the vibrational modes present in the reflectance spectra, we excite the sample so that the electric field is always perpendicular to the c-axis, ensuring to excite only the $E_1$ phonon branch. Additionally, in order to match the light in-plane momentum to the polariton momentum and excite the SPhPs, the attenuated total reflectance (ATR) technique has been used in Otto configuration, with a ZnSe prism ($n_{prism}$ = 2.37). The angle of incidence (θ) at the prism–air gap interface was controlled from 39° to 59°, allowing to scan the polariton dispersion curve through $k_x = k_0 n_{prism} \sin(\theta)$, where $k_0$ is the light momentum in vacuum.

**Numerical models.**

**(i) TMM model with full layer structure.**

To define the structure composed of multiple alternating layers, we account for all individual materials, in this case ZnO and (Zn,Mg)O, each of them defined by its individual dielectric function and thickness. The dielectric functions of the two constituent materials, $\epsilon_{ZnO}$ and $\epsilon_{(Zn,Mg)O}$ are shown in **Equation (2)** and **Equation (3)**, respectively. To obtain the theoretical reflectance spectra of the ZnO/(Zn,Mg)O heterostructure and to fit the model to the experimental measurements to obtain the real parameters of the measured sample, we employ a commercial package based in the transfer matrix method (TMM).

**(ii) TMM with Effective medium theory (EMT).**

To define the metamaterial, the effective medium theory (EMT) for a uniaxial anisotropic structure is used to calculate the in-plane ($\epsilon_\parallel$) and the out-of-plane component ($\epsilon_\perp$) of the effective dielectric function, as a function of the dielectric functions of the two constituent materials. The expressions of $\epsilon_\parallel$ and $\epsilon_\perp$ are shown in **Equation (4)** and **Equation (5)**, respectively.

**(iii) Dispersion curves for high k using EMT.**

The frequency dispersion of the SPhP in the metamaterial was performed by evaluating the imaginary part of the reflection coefficient ($r_p$) for high values of $k_\parallel$, using the TMM for an anisotropic material.[29] In this calculation the metamaterial layer was modeled with the EMT. As shown in Ref.[28], the imaginary part of the reflection coefficient peaks at the polariton resonance frequency allowing to readily observe the polariton frequency dispersion.

**ACKNOWLEDGEMENTS**

This work was partly supported by: Project PID2020-114796RB-C2 funded by MCINN/ AEI /10.13039/501100011033/; Project PDC2023-145827-C2 funded by MICIU/AEI /10.13039/501100011033/ and by European Union Next Generation EU/ PRTR and EU Horizon 2020 Research and Innovation Program under grant agreement No. 665107 (project ZOTERAC); P. I. acknowledges the FPU (Formación de Profesorado Universitario) predoctoral contract from the Spanish Ministry of Science, Innovation, and Universities (MICIU).